\documentclass[aps,prl,twocolumn,showpacs,superscriptaddress]{revtex4-1}
\usepackage{graphicx}
\usepackage{amssymb}
\usepackage{color}
\usepackage{amsmath}
\usepackage{ulem}
\usepackage{mathrsfs}
\usepackage{footmisc}
\usepackage[pdfstartview=FitH,colorlinks=true,linkcolor=blue,citecolor=blue,urlcolor=blue]{hyperref}

\begin{document}
	
\author{Phillip Weinberg}
\affiliation{Department of Physics, Boston University, 590 Commonwealth Avenue, Boston, Massachusetts 02215, USA}

\author{Marek Tylutki}
\affiliation{Department of Applied Physics, Aalto University School of Science, FI-00076 Aalto, Finland}
\affiliation{Faculty of Physics, Warsaw University of Technology, Ulica Koszykowa 75, PL-00662 Warsaw, Poland}

\author{Jami M. R{\"o}nkk{\"o}}
\affiliation{CSC--IT Center for Science, P.O. Box 405, FIN-02101 Espoo, Finland}

\author{Jan Westerholm}
\affiliation{Faculty of Science and Engineering, {\AA}bo Akademi University, Vattenborgsv\"agen 3, FI 20500, {\AA}bo, Finland}

\author{Jan A. {\AA}str\"om}
\affiliation{CSC--IT Center for Science, P.O. Box 405, FIN-02101 Espoo, Finland}

\author{Pekka Manninen}
\affiliation{CSC--IT Center for Science, P.O. Box 405, FIN-02101 Espoo, Finland}

\author{P\"aivi T\"orm\"a}
\affiliation{Department of Applied Physics, Aalto University School of Science, FI-00076 Aalto, Finland}

\author{Anders W. Sandvik}
\affiliation{Department of Physics, Boston University, 590 Commonwealth Avenue, Boston, Massachusetts 02215, USA}
\affiliation{Beijing National Laboratory of Condensed Matter Physics and Institute of Physics, Chinese Academy of Sciences, Beijing 100190, China}

\title{Scaling and diabatic effects in quantum annealing with a D-Wave device}

\begin{abstract}
  We discuss quantum annealing of the two-dimensional transverse-field Ising model on a D-Wave device, encoded on $L \times L$ lattices
  with $L \le 32$. Analyzing the residual energy and deviation from maximal magnetization in the final classical state, we find an optimal $L$ dependent
  annealing rate $v$ for which the two quantities are minimized. The results are well described by a phenomenological model with two powers of $v$ and
  $L$-dependent prefactors to describe the competing effects of reduced quantum fluctuations (for which we see evidence of the Kibble-Zurek mechanism) 
  and increasing noise impact when $v$ is lowered. The same scaling form also describes results of numerical solutions of a transverse-field Ising model 
  with the spins coupled to noise sources. We explain why the optimal annealing time is much longer than the coherence time of the individual qubits.
\end{abstract}

\date{\today}

\maketitle

The  prospect of simulating theoretical quantum many-body Hamiltonians with controllable engineered systems is now an important motivation for atomic and 
quantum device physics \cite{lloyd96,trabesinger12,georgescu14}. Systems explored for creating such ``synthetic quantum matter'' include 
ultracold gases \cite{jaksch05,garc_a_ripoll05,diehl08,bakr09,simon11,bloch12}, photonic devices \cite{aspuru-guzik12,peruzzo14,hartmann16,noh16,c_harris17},
polaritons \cite{berloff17}, and trapped ions \cite{cirac95,james98,porras04,friedenauer08,haffner08,kim10,barreiro11}. Another emerging simulation 
platform is large arrays of superconducting qubits \cite{dwave_systems,r_harris10,r_harris13,dickson13,johnson11,harris18}, which were originally 
envisioned in the context of quantum annealing (QA) as efficient solvers of classical optimization problems mapped to Ising like Hamiltonians
\cite{finnila94,kadowaki98,brooke99,farhi01,roland02,suzuki05,mitchell05,morita08,das08,caneva08,caneva09,heim15,zanca16,knysh16}. To reach
the classical ground state (the problem solution) in a QA process, strong quantum fluctuations are initially induced by applying a transverse 
field, which is quasi-adiabatically reduced to zero. QA devices operating according to this principle have entered industrial production and applications 
beyond the academic setting~\cite{dwave_systems}, motivated by the hope of more efficient solutions of NP-hard problems~\cite{farhi01,lucas14} and, more
recently, quantum enhanced machine learning~\cite{amin18,benedetti16}. It is still unclear what systems (classes of optimization problems) are amenable 
to significant speedups, and to what extent QA can be realized in actual
devices \cite{altshuler10,ronnow14,katzgraber14,amin15,katzgraber15,heim15,mayor15,venturelli15,marshall16,marshall17}.

\begin{figure}[b]
\includegraphics[width=60mm]{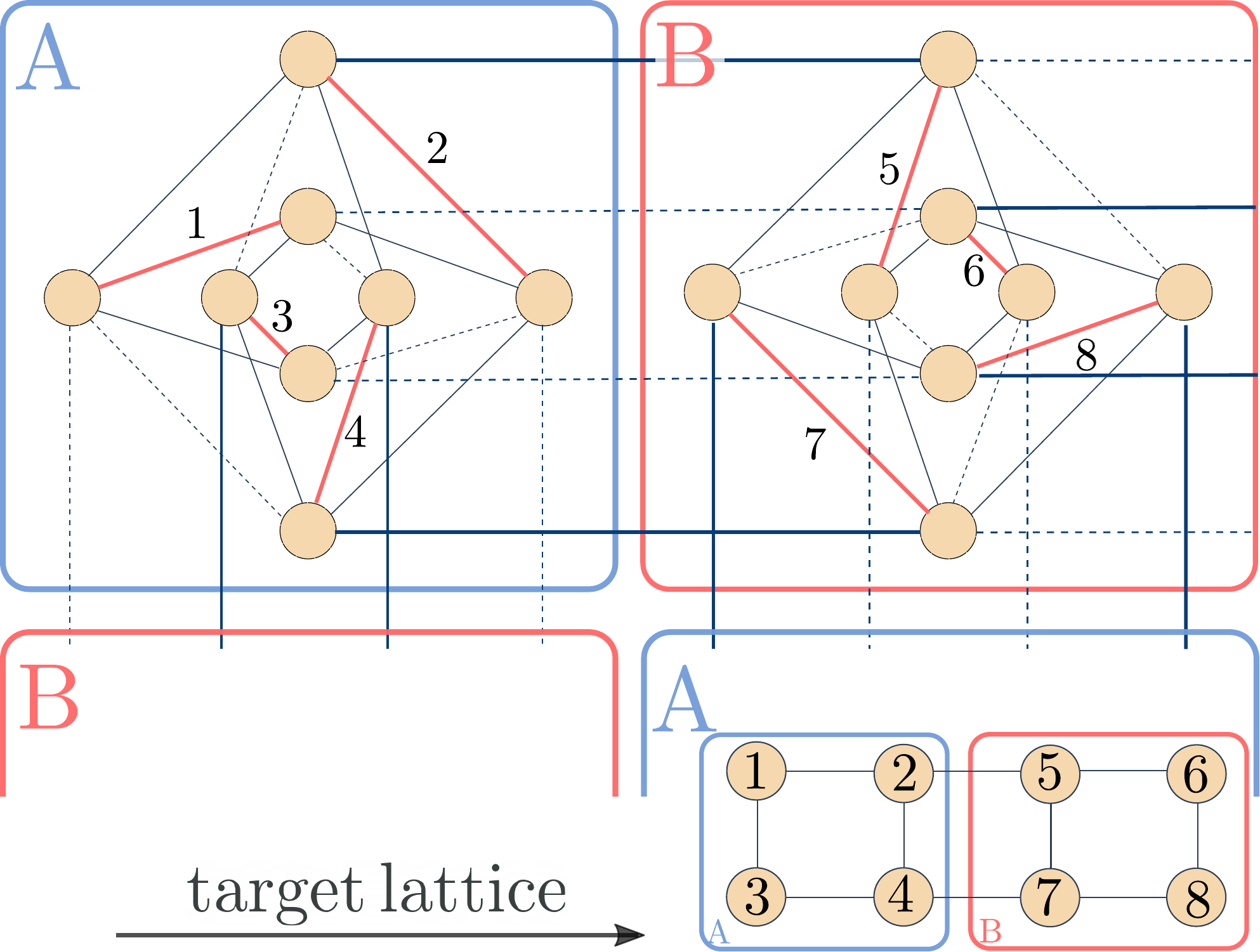}
\caption{Illustration of the DWQ Chimera graph and the embedding of our target open square Ising lattice 
(upper left corner shown). The red links show how the physical qubits are coupled with $J_{\rm HC}$ to
create the logical qubits of the target model. The solid and dashed lines correspond to the active and inactive couplings, 
respectively. The embedding requires two types of Chimera patterns that tile the plane like a chess board. See
SM for details \cite{sm}.}
\label{fig:obc}
\end{figure}

While the question of quantum speedups is essential, the potential of using QA devices as generic quantum many-body
emulators motivates a broader range of investigations into the devices and how they can be exploited for probing various quantum phenomena. As an example, recently
a QA device produced by D-Wave Systems was used in an impressive study of a quantum phase transition of a quantum spin glass \cite{harris18}. An
important question in applications of QA devices, for optimization or quantum simulation, is whether the desired adiabatic evolution is sufficiently realized
in the presence of noise (the environment) and finite annealing time. This question motivates studies of the dependence of measured properties on the annealing
time \cite{mishra18,dickson13,denchev16,chancellor16,gardas18}, which also impacts the effects of noise. For this purpose, it may
be particularly fruitful to implement simple, uniform model Hamiltonians to avoid distractions of not fully understood random couplings \cite{chengwei15}.
Such a study was already carried out with the  one-dimensional transverse-field Ising model (TFIM) coded on a D-Wave device \cite{gardas18}, but the results
did not exhibit any obvious scaling behavior.

In this Letter, we report success of a scaling approach for a two-dimensional (2D) Ising model, with data generated on the D-Wave DW 2000Q$\_2$$\_$1 (DWQ)
solver \cite{dwave_systems}. We observe how the improved adiabaticity with lowered annealing rate competes with diabatic noise mechanisms causing opposite effects,
leading to an optimal annealing rate. We introduce a unified scaling ansatz which can account phenomenologically for both mechanisms in the DWQ
and also describes numerical results for QA of a model Hamiltonian with external noise.

%------------------------------------------------------------------

{\it Model Embedding.}---The DWQ device emulates the TFIM with an array of superconducting loops which form qubits corresponding to 
spin-1/2 operators $\mathbf{\sigma}_i$ (Pauli matrices). The ``Chimera'' interaction graph is made out of cells of eight 
qubits, each connected to six other qubits (five on the graph boundary) and a longitudinal field $h_i$, thus realizing an Ising Hamiltonian
of the form $\mathcal{H}_{\rm class} = \sum_{\langle ij \rangle} J_{ij} \sigma^z_i \sigma^z_j + \sum_i h_i \sigma^z_i$. Here $J_{ij}$ and $h_i$ 
are dimensionless couplings with values in $[-1,1]$. All qubits are coupled to a transverse field, which along with the overall interaction
strength is varied through a time-dependent parameter $s(t) \in [0,1]$ for a total Hamiltonian
\begin{equation}
\mathcal{H}_{\rm TFIM} = A(s) \sum_i \sigma^x_i + B(s) \mathcal{H}_{\rm class}.
\label{htifm}
\end{equation}
$A(s)$ and $B(s)$ are smooth non-linear functions of $s$ \cite{r_harris10,dickson13,harris18} such that $B(0) = 0$ initially and $A(1) = 0$ at the
end of the QA process. Within these bounds there is some flexibility in $s(t)$. The total annealing time can be varied from microseconds to milliseconds. 

For geometries that do not fit on the Chimera graph, logical qubits can be created by coupling two or more physical qubits together with a
``high-cost'' coupling \cite{dwavemanual}, $-J_{\rm HC}=1$, to keep their values mostly the same. The logical qubits can then be coupled
in more complicated geometries \cite{gardas18,harris18,venturelli15}. Here we realize $L\times L$ open-boundary lattices (tiles) 
by using logical qubits constructed from two physical qubits; see Fig.~\ref{fig:obc} and Supplemental Material (SM) 
\cite{sm}. Our target model has equal nearest-neighbor ferromagnetic couplings $J_{ij}=-J_{\rm Ising}$ and $h_i=0$. The DWQ has 2048 qubits, and the maximum 
lattice size for our target model is hence $32 \times 32$. Smaller tiles are implemented by appropriately zeroing some couplings, and for $L \le 16$
we can study several tiles in parallel. The device typically has some nonfunctioning qubits, and we treat all logical qubits affected by defects
as vacancies, completely isolating them by zeroing the corresponding couplings. The fraction of vacancies is typically at most a few percent, and 
tiles with an excessive number of vacancies are not included in the analysis.

We use the maximum high-cost coupling in units of frequency, $J_0 = B(1)J_{\rm HC}/\hbar \approx 2$ GHz \cite{dwavemanual}, 
to set the time units in our plots. Our aim is to study the final-state excitation energy and magnetization as functions 
of the annealing time $T$. To this end, we chose the simplest possible protocol---a linear ramp with $s(t) = t/T = vt J_0$, 
with the dimensionless velocity $v=1/(T J_0)$.

%------------------------------------------------------------------

{\it Phase transition and bath effects.}---The 2D TFIM with Ising coupling $J$ and field $h_x$ undergoes a phase transition 
between ferromagnetic and paramagnetic ground states at $h_x/J\approx 3.04$. Thus, in the DWQ embedded model we expect a phase transition for some value
of $s$ that also depends on $A(s)$ and $B(s)$ in Eq.~(\ref{htifm}). The system will traverse the quantum critical point on its way to the final ordered
ferromagnetic classical state, and this point, where the excitation gap has a size-dependent minimum, is the bottleneck for the system to remain in the 
instantaneous ground state during the entire QA process.

Both classical (stochastic dynamics) and quantum (Hamiltonian dynamics) systems exhibit dynamic scaling in the velocity by which a parameter
changes when passing through a critical point sufficiently slowly. In the neighborhood of the phase transition the exponents are predicted by the
Kibble-Zurek mechanism (KZM) \cite{polkovnikov05,zurek05,dziarmaga10,polkovnikov11} and its generalization as an out-of-equilibrium finite-size scaling (FSS)
ansatz \cite{degrandi10,degrandi11,degrandi13,chengwei13,chengwei15,xu17,xu18,kolodrubetz12}. As an example, the residual Ising energy, defined
as $\Delta_E = \mathcal{H}_{\rm class}-E_{\rm I}$, where $E_{\rm I}$ is the Ising energy in the instantaneous ground state, scales as (in $d$ dimensions
with correlation-length exponent $\nu$ and dynamic exponent $z$) $\Delta_E \sim \xi_{\rm KZ}^{-d} \sim L^d v^{\nu d/(1 + \nu z)}$, where $\xi_{\rm KZ}$ is
the freeze-out length \cite{polkovnikov05,zurek05,dziarmaga10,polkovnikov11}. However, in the long-time limit it has been argued that the Landau-Zener
mechanism (LZM) applies, where the adiabatic evolution is only controlled by the minimum gap $\Delta_L \sim L^{-z}$, giving $\Delta_E \sim L^d v^{1/2z}$
\cite{mitchell05,zurek05,caneva08,caneva09,knysh16,zanca16}. Other types of dynamics, e.g., quantum coarsening,
may also play a role in the long-time limit \cite{Chandran13,Maraga16}.

The KZM and LZM assume an isolated system, but in a device there is always some coupling to a bath or other sources of noise. Works on QA in
open quantum systems have discussed decoherence due to defects produced by the environment at a rate determined by the temperature and the couplings
to the system \cite{dutta16,chenu17,patane08}. If the bath induced defect density remains low throughout the QA process, there may still be a
regime where the scaling depends on the critical point as in the KZM or LZM. However, in some cases the bath can lead to new power 
laws \cite{yan18} or destruction of the critical point \cite{hoyos07,hoyos08}. Decoherence can also some times assist the QA process in approaching
the classical ground state \cite{nishimura16,mishra18,kechedzhi16,arceci17}. Given the desire to better understand and characterize the 
QA process, we will present a systematic FSS analysis of annealing data obtained with the DWQ device.

\begin{figure}[t]
\includegraphics[width=73mm]{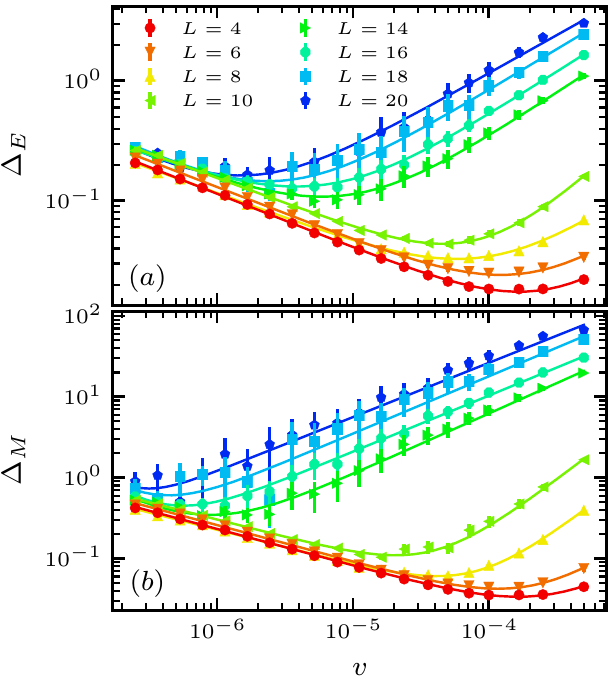}
\caption{Mean values of the excess energy (a) and the magnetization deficit (b) for different lattice sizes after DWQ annealing runs
with $J_{\rm HC}$ set to the maximum possible value \cite{dwavemanual} and $J_{\rm Ising}=0.5J_{\rm HC}$. Each point was calculated using averages of at least
$2 \times 10^4$ independent measurements. The curves are fits to Eq.~(\ref{eq:heating_eq}). The case $L=12$ was not studied.} 
\label{fig:2000Q}
\end{figure}

%------------------------------------------------------------------

{\it Results.}---We investigate the excess Ising energy $\Delta_E$ and the deviation of the magnetization from its maximal (absolute)
value $N$ (the number of qubits), $\Delta_M = N - |\sum_i\sigma^z_i|$.
We saw no significant difference between observables calculated from the logical qubits versus the physical qubits, reflecting the rarity
of violations of the $J_{\rm HC}$ constraint. Here we present results for the physical qubits on the Chimera graph.
In the DWQ device a projective measurement is performed at the end of each annealing run, returning a product
state in the $\sigma^z$ basis. We repeat the annealing protocol at least $2 \times 10^4$ times (over multiple days)
and average over the final configurations.

In Fig.~\ref{fig:2000Q} we show results from the DWQ with $J_{\rm Ising}=0.5$ (see SM \cite{sm} for the motivations for this choice) and lattice sizes
up to $L=20$. We have carried out runs up to $L=32$ (see SM \cite{sm}), but we excluded the larger systems here because of large statistical fluctuations
and no distinct minimums in the accessible velocity window. For the smallest systems, in Fig.~\ref{fig:2000Q} we see that $\Delta_E$ and $\Delta_M$
are already close to their smallest attainable values at the highest $v$, and upon reducing $v$ both quantities increase. Clearer minimums (optimal
velocities) form at lower velocities as $L$ increases. We find power laws emerging on both sides of the minimums. An
optimal annealing rate is consistent with general expectations for QA in a system coupled to a heat bath or noise 
\cite{dutta16,chenu17,patane08,keck17,patane09,nalbach15}, provided that the temperature or noise strength is not too high \cite{arceci18}. 
To our knowledge, the size dependence has not been discussed extensively.

\begin{figure}[t]
\includegraphics[width=70mm]{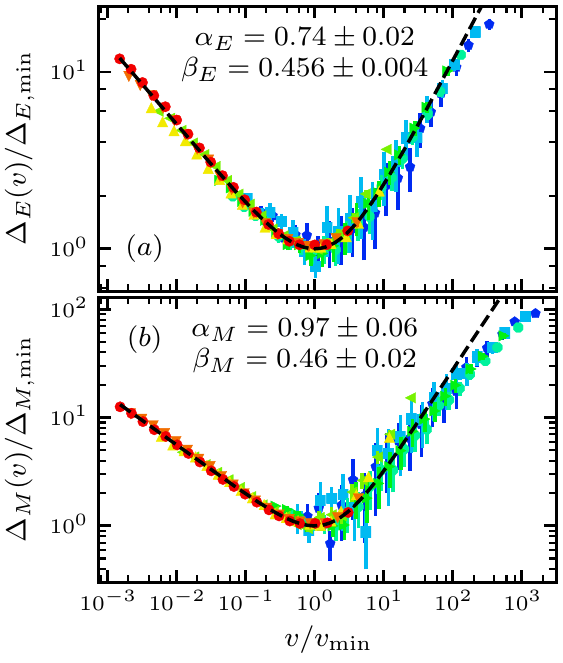}
\caption{Scaling collapse of the data from Fig.~\ref{fig:2000Q}. The curves represent fits to Eq.~(\ref {eq:heating_re}) for $L=10$-$20$,
giving the exponents $\alpha$ and $\beta$ for the two quantities shown in the respective panels.}
\label{fig:2000Q_exp}
\end{figure}

A candidate for a phenomenological model to fit the data is simply a sum of two power laws:
\begin{equation}
f(v)=a_L v^\alpha+b_L v^{-\beta},
\label{eq:heating_eq}
\end{equation}
and $a_L$, $b_L$, $\alpha$, and $\beta$ positive parameters (different for $f=\Delta_E$ and $f=\Delta_M$). The first term accounts for the defect production
from non-adiabatic  QA (which decreases as $v$ decreases), while the second term is the contribution of defects from the
bath (which should increase as $v$ decreases \cite{patane08}). As shown in Fig.~\ref{fig:2000Q}, the form indeed fits all the data. For the larger
systems $a_L$ scales as $L^{2}$ for both the energy and the magnetization (see SM \cite{sm}), which is consistent with both the KZM and LZM
(extensive defect production). The prefactor $b_L$ of the bath term is almost independent of $L$ (as seen in the low-$v$ data in Fig.~\ref{fig:2000Q}
and further analysis in SM \cite{sm}), where one might instead have expected an extensive contribution. This behavior may be an indication of 
highly non-uniform noise (see discussion in SM \cite{sm}) and calls for further investigations of the couplings of the DWQ qubits to the environment.

Even without detailed understanding of the noise, our proposed form (\ref{eq:heating_eq}) provides a way to quantify the competition between adiabatic
and diabatic mechanisms. The optimal values $f_\mathrm{min}(L)$ for both the energy and the magnetization ($f_\mathrm{min}=\Delta_{E,\mathrm{min}}$ or
$f_\mathrm{min}=\Delta_{M,\mathrm{min}}$) and the corresponding velocities $v_\mathrm{min}(L)$ can be used to define rescaled velocities and observables,
\begin{equation}
u \equiv \frac{v}{v_\mathrm{min}},~~~~g(u)\equiv\frac{f(u v_\mathrm{min})}{f_\mathrm{min}}=\frac{\beta u^\alpha+\alpha u^{-\beta}}{\alpha+\beta},
\label{eq:heating_re}
\end{equation}
where the last form follows from Eq.~(\ref{eq:heating_eq}); note the absence of the factors $a_L$ and $b_L$. In Fig.~\ref{fig:2000Q_exp} we show
the rescaled data along with fits to Eq.~(\ref{eq:heating_re}). The resulting exponents $\alpha$ and $\beta$ are displayed in Fig.~\ref{fig:2000Q_exp}.
The QA exponents $\alpha_E$ and $\alpha_M$ agree remarkably well with the Ising KZM forms ($d=2,z=1,\nu\approx 0.630,\beta\approx 0.326)$;
$\Delta_E \sim  v^{d\nu/(1 + \nu)} \sim v^{0.77}$ and $\Delta_M \sim v^{(d\nu+\beta)/(1+\nu)} \sim v^{0.97}$
\cite{degrandi10,degrandi11,degrandi13,chengwei13,chengwei15,xu17,xu18,kolodrubetz12} 
(see SM \cite{sm} for discussion of the exponents).
The LZM energy is $\Delta_E \sim  v^{1/2}$ (and $\Delta_M$ is undetermined). The observed KZM scaling indicates
that the accessible annealing times, before the cross-over to the noise regime, are still not in the long-time limit where other
mechanisms \cite{Chandran13,Maraga16} take over (see also SM \cite{sm}).

{\it Modeling the Bath.}---To understand the diabatic effects responsible for the second term in Eq.~(\ref{eq:heating_eq}), we use a simple model
of decoherence; the TFIM with a noisy transverse field (similar to Refs.~\cite{dutta16,chenu17,smelyanskiy17}). Since calculations for 2D models 
are limited to very small systems, and we also do not intend to describe the details of the DWQ, we use a 1D model to test the proposed generic 
scaling forms in Eqs.~(\ref{eq:heating_eq}) and (\ref{eq:heating_re}).
We find qualitatively similar behaviors with a size-dependent optimal velocity. Note that the KZM predicts universal scaling in the velocity
regime where the noise is not important, with exponents given by the relevant universality class and dimensionality (see SM \cite{sm}).

The Hamiltonian consists of coherent and noisy parts; $\mathcal{H}(t) = \mathcal{H}_0(t)+\mathcal{V}_\mathrm{noise}(t)$, where
\begin{equation}
  \mathcal{H}_0(t) = -\left(\frac{t}{T}\right)^{\hskip-1mm 2} \sum_{i=1}^{L-1} \sigma_i^z\sigma_{i+1}^z
  -\left(1-\frac{t}{T}\right)^{\hskip-1mm 2}\sum_{i=1}^L\sigma_i^x
\label{eq:pure_anneal}.
\end{equation}
Here $T$ is the annealing time and the time dependence is similar to that in the DWQ. 
The noise couples to the transverse field of each qubit
with strength $\lambda$,
\begin{equation}
\mathcal{V}_\mathrm{noise}(t) = \lambda \sum_i\eta_i(t)\sigma_i^x,
\label{eq:noise_model}
\end{equation}
where $\eta_i(t)$ are classical fields representing the interaction with the environment \cite{chenu17}. Experiments run on the DWQ have found 
an approximate $1/\omega^p$ spectrum with $p \approx 0.7$ for the local field ($h_i$) noise \cite{r_harris10,dwavemanual}. The physics is not 
significantly different when the noise is instead added to the transverse field \cite{chenu17}, as we do here. The noise can be summarized with the 
following temporal and spatial
correlations: $\langle\eta_i(t)\eta_j(t')\rangle=\delta_{ij}C(t-t')$, with $C(t-t')$ the autocorrelation function for the noise. We normalize
the noise such that the standard deviation is set to unity and approximate $\eta_i(t)$ as a sum over $10^3$ cosines with frequencies sampled (see details
in SM \cite{sm}) from a power-law spectrum $S(\omega)$ with a cutoff scale $\omega_0$; 
\begin{equation}
S(\omega)=\frac{(\omega/\omega_0)^{-p}e^{-\omega/\omega_0}}{\omega_0\Gamma(1-p)}.
\end{equation}
We set $\omega_0=1$ (given in the natural units of $H_0$), the exponent to $p=0.75$, and noise coupling $\lambda=0.01$.

The simulation starts with the system in the ground state at $t=0$, and the evolution with the Schr\"odinger equation is performed by
a Jordan-Wigner transformation to fermions and solving the Bogoliubov-de Gennes equations \cite{dutta16,zanca16}. To calculate the expectation
value of the energy from the density matrix, we perform many runs with different noise realizations and average over the expectation values 
calculated with the pure state at the end of the run. We did not compute $\Delta_M$, which would be more time consuming.

\begin{figure}[t]
\includegraphics[width=78mm]{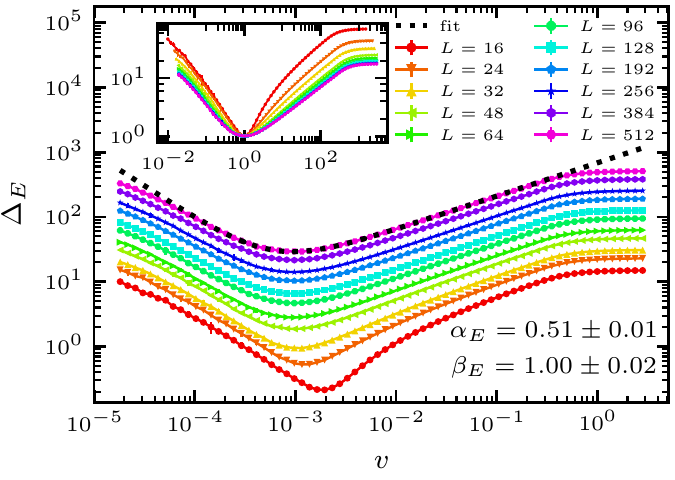}
\caption{Mean residual energy of the 1D TFIM at the end of QA simulations with noise described by Eq.~(\ref{eq:noise_model}) and parameters given in the
text. The inset shows the data rescaled according to Eq.~(\ref{eq:heating_re}). A fit gives the exponents shown in the plot; the $L=512$ form is shown 
as the dashed curve.}
\label{fig:model}
\end{figure}

Figure \ref{fig:model} shows results for various chain lengths. The excess energy first decreases when $v$ is lowered but increases as
$v\rightarrow 0$, similar to the DWQ (Fig.~\ref{fig:2000Q}). The inset shows data collapse with the same kind of rescaling as with the
DWQ data in Fig.~\ref{fig:2000Q_exp}. The prefactors $a_L$ and $b_L$ are both $\propto L$ (see SM \cite{sm}), i.e., the noise effects are extensive
in this case. The KZM and LZM  exponents are identical for this system, $\alpha=1/2$, and $a_L \propto L$, and these power laws agree with the observations.
At high $v$, where the system cannot evolve significantly, the rescaled data approach a constant, corresponding to the properties of the initial state.
Interestingly, in the DWQ data (Fig.~\ref{fig:2000Q_exp}) we also observe similar deviations from the power law at the highest velocities, but there the values 
are still quite far from (about an order of magnitude) those of the ideal fully $x$ polarized initial state.

%------------------------------------------------------------------

{\it Discussion.}---We have shown that QA in the DWQ and a prototypical model system both produce results captured by a simple scaling form,
Eq.~(\ref{eq:heating_eq}), with two power laws describing the competition between quasi-adiabatic annealing and diabatic
effects of a bath. The size-dependent prefactors indicate whether defect production by the two sources is extensive or not, and the powers
of the velocity contain information on the excitation mechanisms at play. Our model system exhibits extensive defect production, as expected,
and the velocity scaling in the annealing regime is consistent with the KZM and LZM (which have the same exponents in the case of the 1D TFIM).
In the DWQ, the velocity scaling is better described by the KZM than the LZM. The bath effects are subextensive, which may indicate highly 
non-uniform effects of the bath \cite{sm}. 

The optimal annealing time, in the DWQ and in the model, is much longer than the coherence time of an individual qubit. As we discuss further in
SM \cite{sm}, correlations among the qubits lessen the impact of noise and lead to a
longer collective time scale of domain ordering. The optimal annealing time should not be seen as a purely quantum mechanical coherence time,
but reflects a fascinating interplay between quantum dynamics and stochastic processes that deserves further study.

Our scaling ansatz should be useful as a generic tool for quantifying QA in the presence of noise sources and baths.
In future experiments with QA devices, it would be interesting to regulate the coupling to the environment in some way, e.g., by changing
the temperature of the system or by introducing additional sources of noise. It will also be useful to implement other uniform and non-uniform
Hamiltonians.

\begin{acknowledgments}
  {\it Acknowledgments.}---We would like to thank Edward (Denny) Dahl, Richard Harris, and Mohammad Amin of D-Wave Systems Inc.~for valuable advice, technical 
  assistance, and insightful comments on the manuscript. We also thank Adolfo del Campo, Anushya Chandran, Jacek  Dziarmaga and Jonathan Wurtz for useful 
  conversations. PW and AWS were supported by the NSF under Grant No.~DMR-1710170 and by 
  a Simons Investigator award. MT and PT were supported by the Academy of Finland under Projects No. 307419, No. 303351, and No. 318987, and by the 
  European Research Council (ERC-2013-AdG-340748-CODE). During the later stages of this work, MT 
  was also supported by the Polish National Science Center (NCN), Contract No. UMO-2017/26/E/ST3/00428. PW would like to thank Department of Applied 
  Physics, Aalto University for hospitality and support during a visit. The numerical calculations were performed on the Shared Computing Cluster 
  administered by Boston University’s Research Computing Services. 
\end{acknowledgments}

\clearpage

\setcounter{page}{1}
\setcounter{equation}{0}
\setcounter{figure}{0}
\setcounter{table}{0}
\renewcommand{\theequation}{S\arabic{equation}}
\renewcommand{\thetable}{S\Roman{table}}
\renewcommand{\thefigure}{S\arabic{figure}}

\onecolumngrid

\begin{center}

{\large Supplemental Material}
\vskip3mm

{\bf\large Scaling and diabatic effects in quantum annealing with a D-Wave device}
\vskip5mm

Phillip Weinberg,$^{1}$ Marek Tylutki,$^{2,3}$ Jami M.~R{\"o}nkk{\"o},$^{4}$ Jan Westerholm,$^{5}$ \\
Jan A.~{\AA}str\"om,$^{4}$ Pekka Manninen,$^{4}$  P\"aivi T\"orm\"a,$^{2}$ and Anders W.~Sandvik$^{1,6}$
\vskip3mm

{\noindent\it\small  
{$^1$Department of Physics, Boston University, 590 Commonwealth Avenue, Boston, Massachusetts 02215, USA}\\
{$^2$Department of Applied Physics, Aalto University School of Science, FI-00076 Aalto, Finland}\\
{$^3$Faculty of Physics, Warsaw University of Technology, Ulica Koszykowa 75, PL-00662 Warsaw, Poland} \\
{$^4$CSC--IT Center for Science, P.O. Box 405, FIN-02101 Espoo, Finland} \\
{$^5$Faculty of Science and Engineering, {\AA}bo Akademi University, Vattenborgsv\"agen 3, FI 20500, {\AA}bo, Finland} \\
{$^6$Beijing National Laboratory of Condensed Matter Physics and Institute of Physics, \\ Chinese Academy of Sciences, Beijing 100190, China}\\
}\vskip8mm

\end{center}

\noindent
In Sec.~1 we provide further details on how the 2D Ising lattice is embedded within the Chimera graph of the DWQ and also
discuss the choice of coupling strengths. In Sec.~2 we discuss the origin of the KZM exponents used to interpret the results in
Figs.~\ref{fig:2000Q_exp} and \ref{fig:model}. In Sec.~3 we present data for larger tile sizes than those discussed 
in the main paper and analyze the size dependence of the prefactors of the velocity powers in the QA-bath scaling ansatz, Eq.~(\ref{eq:heating_eq}).
In Sec.~4 we provide QA results for the 1D TFIM with a noise source coupled to a single spin, complementing the results for
the model with noise sources at all spins in the main text. In Sec.~5 we discuss the coherence time of the noisy spins in the
model and contrast that with the optimal annealing time in the QA process in the presence of a bath. Details of the generation of the noise
signal in the TFIM calculations are provided in Sec.~6.

\subsection{1. Embedding of the Ising square lattice on the D-wave Quantum Processing Unit}

\begin{figure}[b]
	\includegraphics[width=112mm]{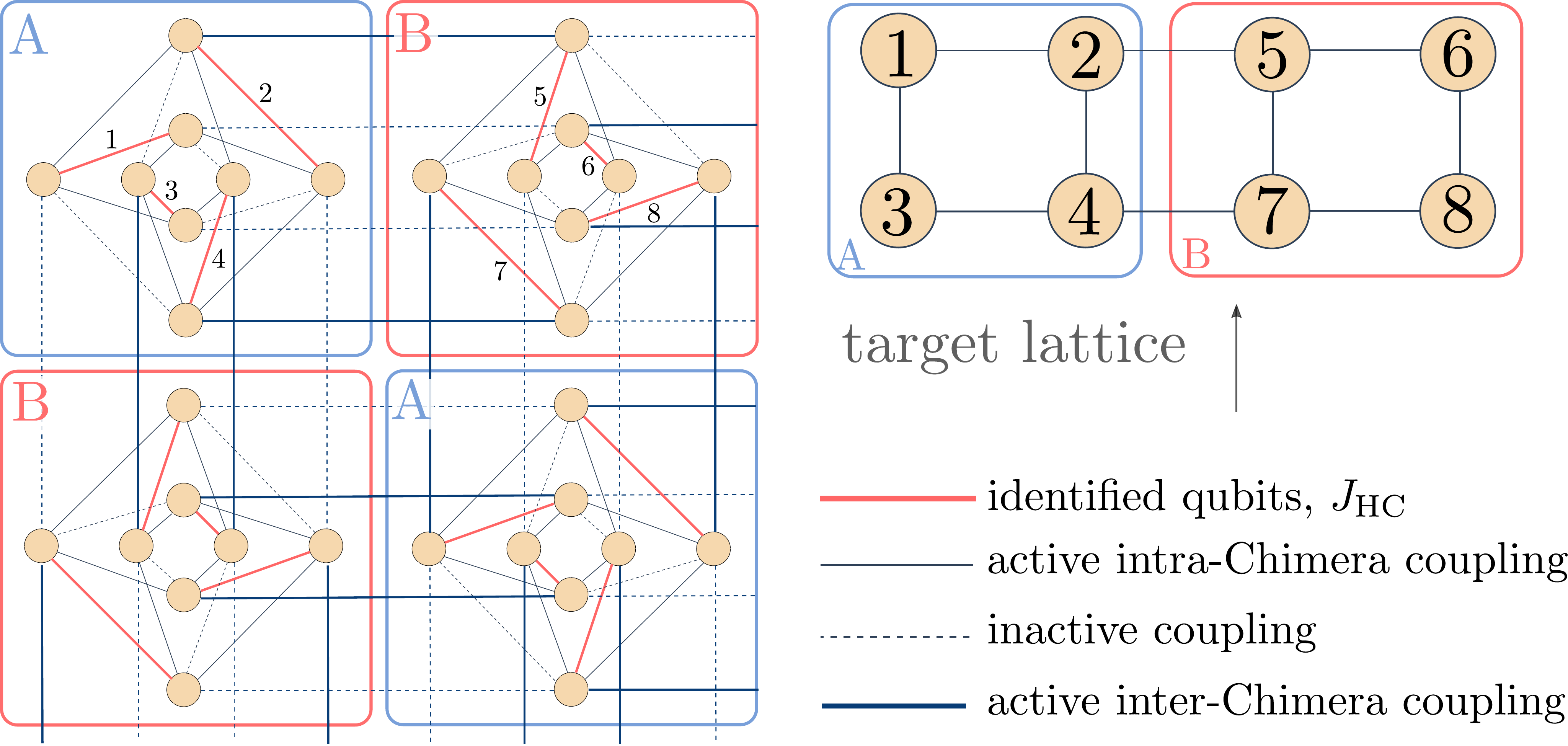}
	\caption{Schematic representing the Chimera geometry and the embedding of our target square lattice with open boundaries
         (upper left corner shown); an extended version of Fig.~\ref{fig:obc} in the main paper. A logical qubit comprises two physical
         qubits coupled by $J_{\rm HC}$ (shown as red lines). In our implementation the nearest-neighbor coupling of strength $J_{\rm Ising}$ 
         is implemented as the sum of two couplings of strength $J_{\rm Ising}/2$ between logical qubits in  the same cell (shown as thinner 
         black lines), while there is only one coupling of strength $J_{\rm Ising}$ between qubits in different cells (shown as thicker black lines).}
	\label{sup:fig:obc}
\end{figure}

We devised an embedding scheme for a square lattice with nearest-neighbor Ising couplings and open boundary conditions To do this, we identify
pairs of physical qubits in each eight-qubit Chimera cell of the D-wave Quantum Processing Unit (QPU) with a single logical qubit on the square lattice.
Using the intra-cell couplings, we strongly couple each of the pairs of physical qubits (labeled 1-8 in the two upper cells in Fig.~\ref{sup:fig:obc})
with the maximum allowed (ferromagnetic) strength $-J_{\rm HC}=1$, and $B(1)J_{\rm HC}/\hbar=2$ GHz in frequency units, to force these qubits to have
the same value. In the discussion we normalize the unit of time such that $B(1)J_{\rm HC}/\hbar=1$. The remaining intra-cell couplings are used to couple the logical
qubits together with their nearest neighbors via the Ising interaction $-J_{\rm Ising}$ with $0 < J_{\rm Ising} < 1$. Thus, we can tile the entire
Chimera graph of the DWQ by selective activation of the inter-cell couplings, setting the unused couplings to zero.  For this embedding to work, we
have to implement two different arrangements of the couplings in the Chimera cells, A and B type. We set them in an alternating pattern across the
lattice; see Fig.~\ref{sup:fig:obc} (an extended version of Fig.~1 of the main paper). The resulting square lattice has $32 \times 32$ sites, but, by
zeroing appropriate inter-tile couplings, we can simulate also smaller tiles of size $L \times L$, and for $L \le 16$ we can obtain more than one independent
tile (assuming the unused couplings really are zero). 

In theory we would like to have $J_{\rm Ising}\ll J_{\rm HC}$; however, for the data we present, we have chosen $J_{\rm Ising}=0.5$. The first reason for
this choice is imperfections in the couplings in the QPU. The typical deviation from the desired coded value is on the order of $0.01$, which sets a
lower bound on how small the couplings can be without becoming too randomized. Note that the value of $J_{\rm Ising}$ also sets the effective scale of the
minimum and maximum annealing times, i.e., in the case of an ideal isolated quantum system the longest allowed QA process would be more adiabatic for
larger $J_{\rm Ising}$. This is another reason for not setting $J_{\rm Ising}$ too small. We did the majority of the runs with $J_{\rm Ising}=0.5$, after
experimentation to obtain the clearest minimums in the observables investigated.

\begin{figure}[t]
	\includegraphics[width=160mm]{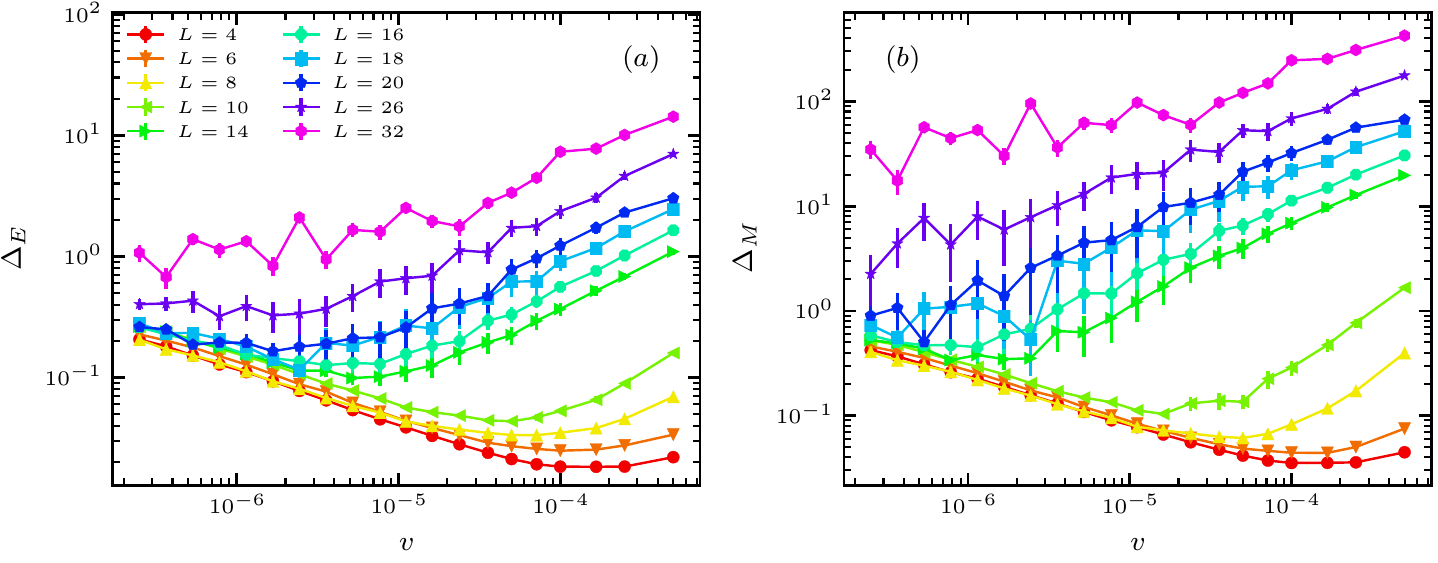}
	\caption{Results from the DWQ device for all system sizes considered.
         Panel $(a)$ shows the mean excess energy defined as the expectation value of the classical
         Hamiltonian over one tile at the end of the annealing minus the energy in the perfectly ordered ground state. Panel $(b)$ shows the sampled
         average absolute value of the magnetization relative to the extreme values $\pm |N|$ of a tile. We have removed contributions from the reported
         broken qubits (nonfunctional qubits or links in the Chimera geometry) for both quantities. Each point was calculated using averages of at
         least $2\times 10^4$ independent annealing runs.}
         \label{sup:fig:dwave_data}
\end{figure}

The DWQ QPU performs the annealing run by quasi-adiabatically turning the transverse field to zero and, simultaneously, the Ising couplings to maximum
strength. At the end of the annealing run the machine performs a projective measurement in the $\sigma^z$ basis. We can therefore measure the Ising energy
and the $z$ magnetization. The measurement returns a product state over the entire QPU lattice, but for system sizes $L\le 16$ the logical system comprises
multiple tiles in a single configuration. Our working assumption is that each of these
tiles is independent of one another, and so when we average the configurations we treat each tile as an independent measurement of the system of the
programmed size $L \times L$. We performed on the order of $2 \times 10^4$ runs per system size, but for the smaller systems the effective number of samples is
multiplied by the number of tiles. The error bars in Fig.~\ref{sup:fig:dwave_data} represent one standard deviation of the mean values. They are computed by
data binning in the way done in Monte Carlo simulations, so that near-normal distributions are obtained.

As mentioned already, there are systematic errors in the couplings. The couplings on the boundaries between the tiles are normally distributed
with a standard deviation $0.01$. These couplings are very small compared to the Ising couplings, and we judge that they do not significantly influence
(correlate) the different tiles. This assumption is also supported by the fact that data for single-tile system sizes, $L>16$, collapse onto
the data sets for multiple-tile sizes $L \le 16$ in our analysis.

Another issue we face is that some of the couplings and qubits on the QPU might not be operational. We take the inactive qubits (which are reported by
the device) into account by treating them
as non-magnetic impurities in the Ising lattice, i.e., for all logical spins containing a broken physical qubit or bond, we set all the couplings to that
logical qubit to zero and do not include such vacancy spins when computing the energy and magnetization. In the device we used, these defects were spread out,
and so each tile had a very low density of defects (at most a few percent, and the rare cases with more defects were discarded when computing averages).

The classical Ising part of the Hamiltonian, Eq.~(\ref{htifm}), has a 'gauge' symmetry which allows us to multiply some spins by a factor $g_i = \pm 1$ and
correspondingly transform all coupling constants $J_{\ij} \to  J_{ij} g_i g_j$ and the longitudinal field (which we do not use here) $h_i \to h_i g_i$.
Such a transformation leaves the spectrum of the full quantum model
invariant as long as there are no shared links $\langle ij\rangle$ among the
transformed spins. The most common transformation of this kind is between the ferromagnet and the antiferromagnet. In the case of the DWQ, performing the
QA runs with a transformed Hamiltonian instead of the original ferromagnet may partially mitigate global field effects that slightly break the spin
inversion symmetry (which would be manifested in different probabilities for reaching the ``up'' and ``down'' magnetized states). We have performed runs
with both ferromagnetic and antiferromagnetic couplings and in each case found only a small inbalance in the ordered states. For the results reported
here, we used ferromagnetic couplings. In principle it might be even better to carry out the transformation for randomly selected spins (with the
constraint of no shared links), using different random transformations for each QA run (gauge averaging \cite{chancellor16})
in the hope of canceling out non-uniform field effects. We have not tested this more complicated procedure.

Regardless of the Hamiltonian used, the initial state of the DWQ is a paramagnetic state,
i.e., a ground state of the Hamiltonian with $s = 0$, when only the transverse field term is active.

\subsection{2. Kibble-Zurek scaling in the ordered phase}

Here we discuss the origin of the KZM exponents used to interpret the DWQ results in Fig.~\ref{fig:2000Q_exp} and the numerical results in Fig.~\ref{fig:model}.
Both $\Delta_E$ and $\Delta_M$ represent deviations from the respective instantaneous ground state equilibrium values and can in principle be defined for all
values of the control parameter $s$, even though we here have considered only the behavior at the final, classical point $s=1$ of the QA protocol. Thus, for
any value of $s$, by definition $\Delta_E(s)$ and $\Delta_M(s)$ vanish for all $s$ when $v \to 0$, provided that we consider an isolated quantum system.
The assumption of an isolated quantum system should be valid for velocities above the optimal velocity.

While the excess Ising energy $\Delta_E$ is an often studied quantity in the KZM literature, the magnetization $M$ close to the critical point
is normally considered as is, without taking a difference, because $M$ vanishes in the thermodynamic limit at the quantum critical point for $v=0$
\cite{polkovnikov05,zurek05,dziarmaga10,polkovnikov11,degrandi10,degrandi11,degrandi13,chengwei13}. Here we are in a regime where 
$M$ is not small, and we have instead chosen to analyze the small deviation $\Delta_M$.

$\Delta_E$ and $\Delta_M$ can be studied on equal footing
within the KZM framework by using the appropriate critical exponents.
We use $\kappa$ to denote a generic critical exponent for a quantity $X$, i.e., its value in the thermodynamic limit scales as $X/L^d \propto \delta^\kappa$
($d$ being the spatial dimensionality) in the ordered phase, where $\delta=s-s_c$ is the distance to the quantum critical point. The KZM ansatz is then
\begin{equation}
\Delta_X = L^{-\kappa/\nu}f(vL^{z+1/\nu}),
\label{deltaxscaling}
\end{equation}
where the factor $L^{-\kappa/\nu}$ sets the overall magnitude, which should scale in the same way as $X$ itself. The exponent on $L$ originates from the
standard finite-size scaling procedure where $\delta$ is expressed using the correlation length, $\xi \propto \delta^{-\nu}$, which is then replaced by the
system length $L$. Note that $\Delta_X$ is an extensive quantity, $\Delta_X(s)=X(s,v)-X(s,0)$, but there is no factor $L^d$ in the scaling form
(\ref{deltaxscaling}). As we will see below, the extensive behavior can still follow from Eq.~(\ref{deltaxscaling}) in the appropriate velocity
regime. We refer to Refs.~\cite{polkovnikov05,zurek05,dziarmaga10,polkovnikov11,degrandi10,degrandi11,degrandi13} for derivations and tests of
scaling forms of the type in Eq.~(\ref{deltaxscaling}) by adiabatic perturbation theory and other approaches. Here we just summarize the behaviors
in different velocity regimes.

The scaling function $f(vL^{z+1/\nu})$ can be series expanded in the limit $v \ll L^{-(z+1/\nu)}$, which gives $\Delta_X \sim vL^{z+(1-\kappa)/\nu}$ (since
there can be no constant term by definition). In the opposite limit $v \gg L^{-(z+1/\nu)}$ (but $v$ still lower than some $L$ independent value of order
one when expressed in appropriate dimensionless units), the scaling function must reduce to a power law, $f(vL^{z+1/\nu}) \to (vL^{z+1/\nu})^\alpha$,
which is what is normally referred to as Kibble-Zurek scaling. Demanding that the quantity is extensive, $\Delta_X \propto L^d$ (corresponding to
an extensive number of excited defects), the exponent $\alpha$ is uniquely determined to be
\begin{equation}
\alpha=\frac{d+\kappa/\nu}{z+1/\nu}.
\label{cexponent}
\end{equation}
In the case of the magnetization, for the TFIM chain (which exhibits 2D classical Ising universality) $\kappa_M=\beta_{\rm 2D} = 1/8$ while in 2D (3D Ising
universality) $\kappa_M=\beta_{\rm 3D} \approx 0.326$. The Ising energy has a regular contribution (i.e., $E/L^d$ is finite at the critical point in the
ground state) and $\kappa_E=0$ in both 1D and 2D. Note that the exponent $\alpha_E$ still depends on the dimensionality $d$ in Eq.~(\ref{cexponent}).

In our analysis of both the 2D DWQ data and 1D numerical results, we found extensive behaviors (see also Secs.~3 and 4 below), demonstrating that the systems
are in the KZM scaling regime where $v \gg L^{-(z+1/\nu)}$. Moreover, the scaling exponents $\alpha_M$ and $\alpha_E$ were found to be in excellent agreement
with the predicted values given by Eq.~(\ref{cexponent}). It is perhaps surprising that these KZM scaling forms hold even in the classical limit $s=1$, which
is not very close to the critical point. However, since the system does not reach perfect order, it is also effectively closer to the critical point than what
just the final $s$ value suggests. It is also interesting to note the cross-over into the high-velocity limit in the model calculation in Fig.~4 in the main
text; here the observable approaches a constant. Hints of such a cross-over can also be seen in the DWQ results in Fig.~3.

\subsection{3. Additional finite-size scaling results}
           
Figure \ref{sup:fig:dwave_data} shows all of our energy and magnetization results, including the data shown in Fig.~\ref{fig:2000Q} in the main paper and
also additional $L=26$ and $L=32$ results not shown there. As mentioned, for those largest system sizes the fluctuations are larger, and,
for unknown reasons, the error bars for $L=32$ are clearly underestimated. The large fluctuations for some velocity values in this case,
beyond what is expected given the size of the error bars, may indicate some anomalous time dependence of the couplings or local fields, but
under such a scenario it is not clear why similar effects are not present in the smaller tiles. It is possible that the $L=18-26$ tiles do
not cover the putative anomalous region, and for the smaller sizes, where there are multiple tiles present, none or only one of the tiles
may be affected. In any case, also the $L=32$ results do show the same trends as the other data sets.

\begin{figure}[t]
	\includegraphics[width=140mm]{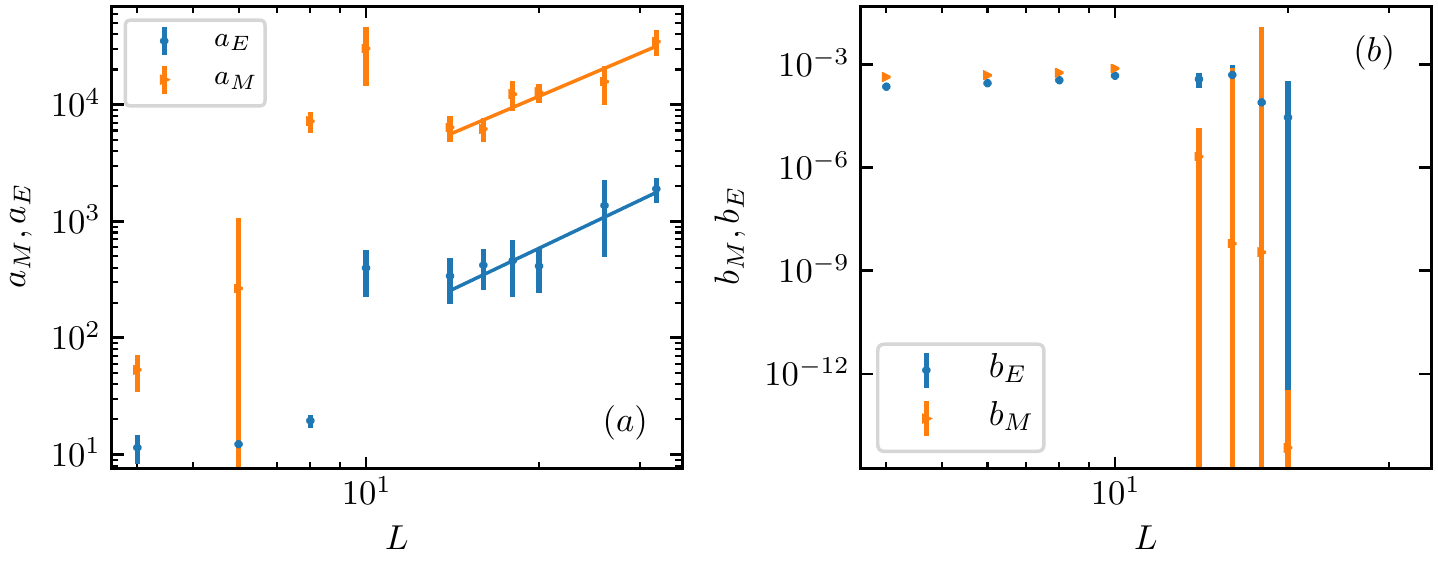}
	\caption{The size-dependent factors $a(L)$ [panel (a)] and $b(L)$ [panel (b)] vs $L$ from fits of the DWQ data with
          Eq.~(\ref{sup:eq:heating_eq}), shown on log-log scales. The blue and orange symbols represent, respectively, the values from the energy
          and the magnetization. The solid lines in (a) represent best-fit power laws $a_E(L) \sim L^{\gamma_E}$ and $a_M(L) \sim L^{\gamma_M}$
          for system sizes $14$ through $32$. For the energy, the exponent is $\gamma_E=2.3 \pm 0.6$ while for the magnetization $\gamma_M = 2.1 \pm 0.4$.
          In (b), we did not perform any fits but both data sets are observed to be almost size independent, i.e.,
          $b_E(L) \sim L^{\mu_E}$ and $b_M(L) \sim L^{\mu_M}$ with exponents $\mu_E,\mu_M\approx 0$.}
          \label{sup:fig:a_fits}
\end{figure}

As discussed in the main text, we fit the data with a model that phenomenologically describes the effects of defects generated by the annealing process
as a sum of two power-laws, Eq.~(\ref{eq:heating_eq}), with one term representing the excitations due to the QA at finite velocity while the other term 
captures the effects of defects produced by the couplings to the bath (which may have multiple components). We repeat the form here for convenience;
\begin{equation}
f=a(L)v^{\alpha}+b(L)v^{-\beta},
\label{sup:eq:heating_eq}
\end{equation}
with $f=\Delta_E$ or $f=\Delta_M$ and different parameters for the two cases. The size independent exponents $\alpha$ and $\beta$ for both the energy
and magnetization were extracted in the main paper in the way illustrated in Fig.~\ref{fig:2000Q_exp}. Here we discuss the scaling of the size dependent
prefactors $a_f(L)$ and $b_f(L)$, $f=E,M$, testing power laws; $a_f(L) \sim L^{\gamma_f}$  and $b_f(L) \sim L^{\mu_f}$. For both the energy and the
magnetization, the values of $a(L)$ obtained from fitting the data in Fig.~\ref{sup:fig:dwave_data} to the form in Eq.~(\ref{sup:eq:heating_eq}) have
large variations versus the system size and large error bars. However for tile sizes $L>10$ we find that the coefficients $a_f(L)$ for both
quantities indeed scale as approximately $L^2$, i.e., extensively, as shown Fig.~\ref{sup:fig:a_fits}(a). Here we have used also the data for $L=26$
and $L=32$, and for reasons discussed above the error bar for $L=32$ is likely underestimated.

In (b) we show the results for the coefficients $b(L)$ arising from the bath contributions. The simplest expectation here is that also these
contributions should be extensive, scaling as $L^{2}$ as was found for $a(L)$ above. However, instead we find essentially size independent behaviors 
for both quantities. Here we have not included the $L=26$ and $L=32$ data, because the uncertainties of the bath terms are too large when no clear minimums 
can be discerned for these system sizes (see Fig.~\ref{sup:fig:dwave_data}).

\subsection{4. Subextensive noise effect in the low-velocity regime}
\label{supp:singlenoise}

\begin{figure}[t]
	\includegraphics[width=160mm]{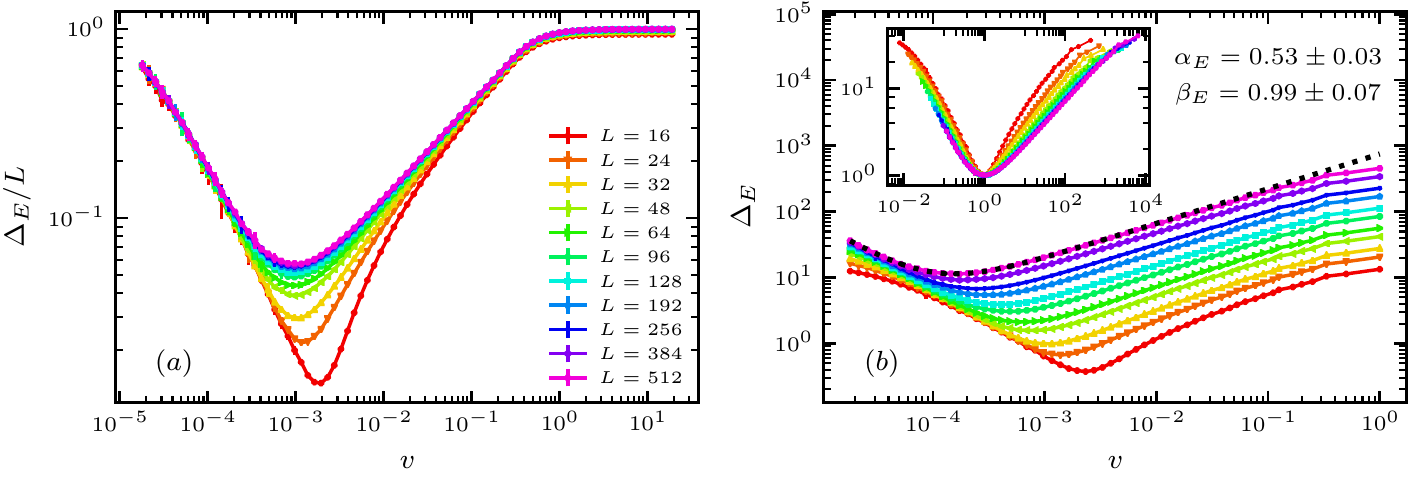}
	\caption{$(a)$ The TFIM data from Fig.~\ref{fig:model} in the main paper with the energy rescaled by $1/L$, to demonstrate the extensive
	  property over the entire velocity range for large system sizes. $(b)$ Results for the TFIM with local (single-source) bath coupling,
          Eq.~(\ref{etalocal}), with the inset showing the same scaling analysis as in Fig.~\ref{fig:model} in the main text. The procedure gave
          the exponents $\alpha_E$ and $\beta_E$ displayed inside the graph. The dotted curve shows the fit to Eq.~(\ref{sup:eq:heating_eq}) for the
          largest system size ($L=512$). The data sets for the different system sizes in (a) and (b) are color coded in the same way according
          to the legends in (a). }
	\label{sup:fig:re_model}
\end{figure}

In the main text and the further analysis above in Sec.~2, we inferred that the defect production in the low-$v$ annealing regime of the DWQ showed a very weak
dependence on the system size. In contrast, the defect production in the 1D TFIM with each spin coupled to a noise source, Eq.~(\ref{eq:noise_model}), showed a much
stronger size dependence, as reflected in the results for the excess energy in Fig.~\ref{fig:model}. In Fig.~\ref{sup:fig:re_model}(a) we scale the same TFIM data as
in Fig.~\ref{fig:model} in a slightly different way, just dividing $\Delta_E$ by $L$ to demonstrate the expected extensive behavior in the whole velocity regime
(for sufficiently large system sizes). The extensive defect production by the model bath is clearly due to the fact that each spin is coupled to an independent
noise source. Conversely, the apparently subextensive defect density in the low-$v$ regime in the DWQ (as reflected by the size independence in the low-$v$ regime
in Fig.~\ref{fig:2000Q}) points to some kind of highly non-uniform effect of the environment, with the noise level (or the impact of the noise) being
much higher on a small number of qubits.

While there may be several sources of noise in the DWQ and a detailed understanding of a potential non-uniformity is lacking, there are some natural
candidates: 1) Qubits at corners (and to a lesser extent elsewhere on edges) of the logical Ising lattice could be more susceptible to noise, 2) Likewise,
corners and edges of the physical Chimera graph may be more susceptible, and 3) qubits close to defective links or qubits may be more noisy. The corners could
clearly give a non-extensive, size-independent effect. Since the defects are sparse and not distributed uniformly, they may also effectively give
rise subextensive contributions.

We would like to test whether having a non-extensive number of decoherent qubits is sufficient to reproduce the results we observe on the DWQ. We can accomplish this
in an extreme way in the 1D TFIM by coupling the system to a noise source at only one part of the chain, here the qubit at one of the edges;
\begin{equation}
\mathcal{V}_\mathrm{noise}(t) = \lambda \eta_1(t)\sigma_1^x.
\label{etalocal}
\end{equation}
We perform the same analysis as in Fig.~\ref{fig:model} in the main text for this new model and present the results in Fig.~\ref{sup:fig:re_model}(b).
Visually the results in the range over which we see minimums in $\Delta_E$ are qualitatively similar to the results from the DWQ device (Fig.~\ref{fig:2000Q}),
however the power-law exponent $\beta$ describing the velocity dependence when the data are scaled according to Eqs.~(\ref{eq:heating_eq}) and (\ref{eq:heating_re})
[shown in the inset of Fig.~\ref{sup:fig:re_model}(b)] is still the same as in the model with noise sources at all qubits (Fig.~\ref{fig:model}). The competition
between the subextensive (size independent) defect production by the noise and the extensive non-adiabatic QA defect production leads to the optimal annealing
time shifting to lower velocities with increasing system size, in a very similar manner as in the DWQ data in Fig.~\ref{fig:2000Q}.

An interesting side note is that there is no clear theoretical argument as to why the defect production from the bath in the DWQ behaves as $\approx\sqrt{v}$
(exponent $\beta \approx 0.5$ in Fig.~\ref{fig:2000Q_exp}). Though we do not know the details of the sources of decoherence in the DWQ device, we can get a hint
of what might be happening by converting velocity into time, e.g. $1/\sqrt{v}\sim\sqrt{t}$. This power-law behavior is reminiscent of diffusion. One might speculate
that this scaling would imply that the energy absorbed by the corners (or defects) is moving into the bulk diffusively as opposed to ballistically, which would
have a natural scale $t$ or $1/v$. This is plausible, as we know that the clean 1D TFIM has ballistic heat transport at low temperatures \cite{sun10} and this
would be a natural explanation for the exponent $\beta\approx 1$ in our model (likely $\beta=1$ exactly). The heat transport properties of the 2D TFIM are
largely unknown.

\subsection{5. Coherence time of a single model qubit}
\label{supp:coherence}

In the main text we discussed how the optimal annealing rate is related to the interplay between the QA process with the uniform transverse field and the effects
of the couplings to the noisy environment. The optimal annealing rate $v_{\rm min}$, defined by the minimums in $\Delta_E$ or $\Delta_M$, was apparent in the data
sets for both the DWQ device and the TFIM. One might naively think that the corresponding time scale $v_{\rm min}^{-1}$ is of the same order of magnitude as the
coherence time of an individual qubit; however this is not necessarily the case as $v_{\rm min}^{-1}$ is a collective time scale. Here we analyze the decoherence
process and extract the coherence time of our model spins. We discuss why the collective time scale in a many-body system can be much longer than the
single-spin coherence time.

\begin{figure}[t]
\includegraphics[width=7.62cm]{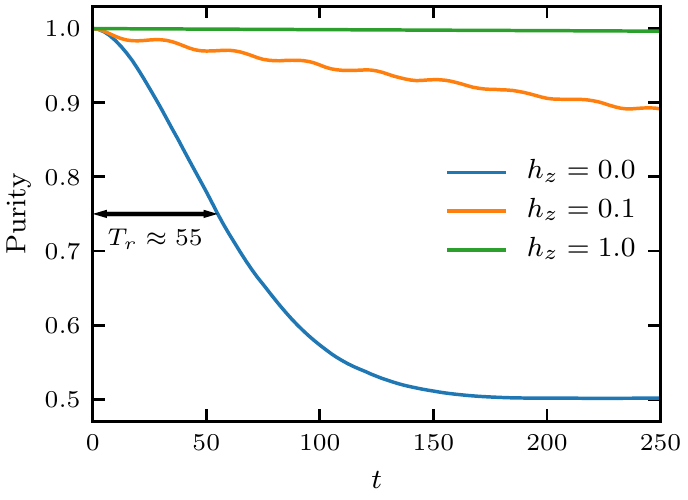}
\caption{Purity, ${\rm Tr}\{\rho^2\}$, of a single qubit evolved after being prepared in the $\sigma^z=1$ state initially. The spin
is coupled to noise and a magnetic field according to Eq.~(\ref{hxzsupp}); results for three different values of the field are shown.
The arrows represent the span of time used here to define the spin-relaxation time of the qubit with no external magnetic field. Here the
time is given in natural dimensionless units in the same way as discussed in the main text. Each curve was produced by evolving the model 
over $10^3$ independent realizations of noise.}
\label{sup:fig:T_r}
\end{figure}

In our model, the noise describes an environment which is creating decoherence via spin flips. For this kind of bath we can define a coherence time, $T_r$,
for a single qubit as the time it takes for the purity of the density matrix to drop from $1$ to $3/4$ (the minimum value being $1/2$) after being prepared
in an eigenstate of $\sigma_z$. Recall that the model of the bath is calculated from an ensemble of pure states generated with a stochastic Schr\"odinger
equation. For a given noise realization the bath is included through an operator which has a stochastic coupling (given by a continuous but highly fluctuating
function). For a single qubit this operator would be
$\mathcal{V}_\mathrm{noise}(t)=\lambda\eta(t)\sigma_x$, where $\eta(t)$ is defined in the same way as the main text (and how the noise is generated in
practice is further discussed below in Sec.~5). We will also consider the effect of adding a magnetic field,
\begin{equation}
	H(t)=h_z\sigma_z+\lambda\eta(t)\sigma_x,
\label{hxzsupp}
\end{equation}
for reasons to be explained below.

Expectation values of observables are calculated as averages over the ensemble of pure states. In an alternate formulation, we can calculate the density
matrix of the qubit and take the average of the density matrices in each noise realization \cite{chenu17}:
\begin{equation}
\rho(t) = \frac{1}{N_r}\sum_r |\psi_r(t)\rangle\langle\psi_r(t)|,
\end{equation}
where $|\psi_r(t)\rangle$ for given realization $r$ is the state evolved with the noisy Hamiltonian and $N_r$ is the number of realizations. Results for
the purity, i.e., ${\rm Tr}\{\rho^2\}$, versus time are displayed in Fig.~\ref{sup:fig:T_r}. The blue curve shows results for the initial state evolved
with just the bath, i.e. $h_z=0$ in Eq.~(\ref{hxzsupp})
(and the noise parameters are the same as used in all other cases). In this case the coherence time $T_r\approx 55$,
which translates to a velocity $v\approx 0.02$. This value is much larger than any of the optimal annealing rates in both the single-spin bath
coupling discussed in the previous section and the model where all qubits are coupled to individual noise sources discussed in the main text. In the
case of the subextensive defect production by the bath (observed in the DWQ as well as in the model in Sec.~3), the effective collective decoherence time
scale even diverges.

An intuitive way to understand why there is a difference in time scales is to turn on the magnetic field in the Hamiltonian, Eq.~(\ref{hxzsupp}), and again
observe the purity versus time. Some results are shown as the green and orange curve in Fig.~\ref{sup:fig:T_r}. As the local magnetic field increases, the
relaxation time $T_r$ increases as well. This is not very surprising, however it does suggest that the collective, apparent coherence time of the system
of coupled qubits can be longer than that of a single qubit due to local, effective magnetic fields acting on each qubit due to the couplings to neighboring
qubits in which order has formed on some length scale.

Our model of decoherence used here is incomplete, in the sence that it only accounts for noise-induced spin flips causing spin relaxation (the relaxation
time normally called $T_1$). It does not involve dephasing (quantified by $T_2$). The time scale reflected in the optimal annealing rate should also not
be taken as a purely quantum mechanical coherence time, but is more reflective of the inability of the statistical noise to destroy the classical correlated
state emerging at the latter stages of the open-system QA process. This intricate phenomenon, originating from a combination of quantum dynamics and
stochastic dynamics, deserves further study.

\subsection{6. Generating the noise signals}
\label{supp:noise}

In the noise terms of the 1D TFIM Hamiltonian in Eqs.~(\ref{eq:noise_model}) and (\ref{etalocal}), as well as the single-spin model defined in Eq.~(\ref{hxzsupp}),
we model the signal $\eta(t)$ at a given site (we here suppress the site index) as a sum of harmonic oscillators at random frequencies and initial conditions:
\begin{equation}
\eta(t)=\frac{1}{\sqrt{N_m}}\sum_{i=1}^{N_m}\bigl (x_i \cos(\omega_i t)+p_i\sin(\omega_i t) \bigr).
\end{equation}
We choose $x_i$ and $p_i$ to be normally distributed with a mean of $0$ and a standard deviation of $1$. As $x_i$ and $p_i$ are normally distributed, this
implies that for any $\omega_i$ and for all times $t$, $\eta(t)$ is also normally distributed. The factor of $1/\sqrt{N_m}$ normalizes $\eta$
so that it has a standard deviation of $1$. The non-equal time correlation function averaged over realizations of $x_i$ and $p_i$, keeping $\omega_i$ fixed,
is given by:
\begin{equation}
\langle\eta(t)\eta(t')\rangle =\frac{1}{N_m}\sum_{i}\cos\bigl (\omega_i (t-t')\bigr ).
\end{equation}
Suppose that we choose $\omega_i$ randomly from a distribution $P\left(\omega\right)$. Then, if we take the limit $N_m\rightarrow\infty$, we can approximate
that sum with the integral over the distribution:
\begin{equation}
\langle\eta(t)\eta(t')\rangle \approx\int d\omega P(\omega)\cos\bigl (\omega (t-t')\bigr) =\mathrm{Re}\bigl(\varphi_P(t-t')\bigr),
\end{equation}
where $\varphi_P(t)$ is the characteristic function of the probability distribution $P(\omega)$. In this way, we can reverse engineer the unequal time correlation
function to obtain the probability distribution for $\omega$. The one issue with this approach is that it is computationally expensive to evaluate this function
in the limit $N_m\rightarrow\infty$, but this is a problem with all methods of sampling a correlated noise signal. However, in the simulations we performed,
we did not see any significant quantitative difference in the scaling analysis when simulating the noise with $1000$ modes versus $100$ modes.  


\vskip-6mm

\begin{thebibliography}{99}

\bibitem{lloyd96}
S. Lloyd, Science {\bf 273}, 1073 (1996).

\bibitem{trabesinger12}
A. Trabesinger, Nat. Phys. {\bf 8}, 263 (2012).

\bibitem{georgescu14}
I. M. Georgescu, S. Ashhab, and F. Nori, Rev. Mod. Phys. {\bf 86}, 153 (2014).

\bibitem{jaksch05}
D. Jaksch and P. Zoller, Ann. Phys. {\bf 315}, 52 (2005).

\bibitem{garc_a_ripoll05}
J. J. Garcı\'ia-Ripoll, P. Zoller, and J. I. Cirac, J. Phys. B. {\bf 38}, S567 (2005).

\bibitem{diehl08}
S. Diehl, A. Micheli, A. Kantian, B. Kraus, H. P. B\"uchler, and P. Zoller, Nat. Phys. {\bf 4}, 878 (2008).

\bibitem{bakr09}
W. S. Bakr, J. I. Gillen, A. Peng, S. F\"olling, and M. Greiner, Nature {\bf 462}, 74 (2009).

\bibitem{simon11}
J. Simon, W. S. Bakr, R. Ma, M. E. Tai, P. M. Preiss, and M. Greiner, Nature {\bf 472}, 307 (2011).

\bibitem{bloch12}
I. Bloch, J. Dalibard, and S. Nascimb\`ene, Nat. Phys. {\bf 8}, 267 (2012).

\bibitem{aspuru-guzik12}
A. Aspuru-Guzik and P. Walther, Nat. Phys. {\bf 8}, 285 (2012).

\bibitem{peruzzo14}
A. Peruzzo, J. McClean, P. Shadbolt, M.-H. Yung, X.-Q. Zhou, P. J. Love, A. Aspuru-Guzik, and J. L. O’Brien, Nat. Commun. {\bf 5}, 4213 (2014).

\bibitem{hartmann16}
M. J. Hartmann, J. Opt. {\bf 18}, 104005 (2016).

\bibitem{noh16}
C. Noh and D. G. Angelakis, Rep. Prog. Phys. {\bf 80} 016401 (2017).

\bibitem{c_harris17}
N. C. Harris, G. R. Steinbrecher, M. Prabhu, Y. Lahini, J. Mower, D. Bunandar, C. Chen, F. N. C. Wong, T. Baehr-Jones, M. Hochberg, S. Lloyd, and D. Englund,
Nat. Photonics {\bf 11}, 447 (2017).

\bibitem{berloff17}
N. G. Berloff, M. Silva, K. Kalinin, A. Askitopoulos, J. D. T\"opfer, P. Cilibrizzi, W. Langbein, and P. G. Lagoudakis,
Nat. Mater. {\bf 16}, 1120 (2017).

\bibitem{cirac95}
J. I. Cirac and P. Zoller, Phys. Rev. Lett. {\bf 74}, 4091 (1995).

\bibitem{james98}
D. James, Appl. Phys. B {\bf 66}, 181 (1998).

\bibitem{porras04}
D. Porras and J. I. Cirac, Phys. Rev. Lett. {\bf 92}, 207901 (2004).

\bibitem{friedenauer08}
A. Friedenauer, H. Schmitz, J. T. Glueckert, D. Porras, and T. Schaetz, Nat. Phys. {\bf 4}, 757 (2008).

\bibitem{haffner08}
H. H\"affner, C. Roos, and R. Blatt, Phys. Rep. {\bf 469}, 155 (2008).

\bibitem{kim10}
K. Kim, M.-S. Chang, S. Korenblit, R. Islam, E. E. Edwards, J. K. Freericks, G.-D. Lin, L.-M. Duan, and C. Monroe, Nature {\bf 465}, 590 (2010).

\bibitem{barreiro11}
J. T. Barreiro, M. Müller, P. Schindler, D. Nigg, T. Monz, M. Chwalla, M. Hennrich, C. F. Roos, P. Zoller, and R. Blatt, Nature {\bf 470}, 486 (2011).

\bibitem{dwave_systems}
The D-wave DW-2000Q quantum annealing device; {\tt  http://www.dwavesys.com/}

\bibitem{r_harris10}
R. Harris {\it et al.},
Phys. Rev. B {\bf 82}, 024511 (2010).

\bibitem{r_harris13}
R. Harris {\it et al.},
Phys. Rev. B {\bf 81}, 134510 (2010).

\bibitem{dickson13}
N. G. Dickson {\it et al.},
Nat. Commun. {\bf 4}, 1903 (2013).

\bibitem{johnson11}
M. W. Johnson {\it et al.},
Nature {\bf 473}, 194 (2011).

\bibitem{harris18}
R. Harris {\it et al.},
Science {\bf 361}, 162 (2018).

\bibitem{finnila94}
A. Finnila, M. Gomez, C. Sebenik, C. Stenson, and J. Doll, Chem. Phys. Lett. {\bf 219}, 343 (1994).

\bibitem{kadowaki98}
T. Kadowaki and H. Nishimori, Phys. Rev. E {\bf 58}, 5355 (1998).

\bibitem{brooke99}
J. Brooke, D. Bitko, T. F., Rosenbaum, and G. Aeppli, Science {\bf 284}, 779 (1999).

\bibitem{farhi01}
E. Farhi, J. Goldstone, S. Gutmann, J. Lapan, A. Lundgren, and D. Preda, Science {\bf 292}, 472 (2001).

\bibitem{roland02}
J. Roland and N. J. Cerf, Phys. Rev. A {\bf 65}, 042308 (2002).

\bibitem{suzuki05}
S. Suzuki and M. Okada, J. Phys. Soc. Jpn. {\bf 74}, 1649 (2005).

\bibitem{mitchell05}
D. R. Mitchell, C. Adami, W. Lue, and C. P. Williams, Phys. Rev. A {\bf 71}, 052324 (2005).

\bibitem{morita08}
S. Morita and H. Nishimori, J. Math. Phys. {\bf 49}, 125210 (2008).

\bibitem{das08}
A. Das and B. K. Chakrabarti, Rev. Mod. Phys. {\bf 80}, 1061 (2008).

\bibitem{caneva08}
T. Caneva, R. Fazio, and G. E. Santoro, Phys. Rev. B {\bf 78}, 104426 (2008).

\bibitem{caneva09}
T. Caneva, R. Fazio, and G. E. Santoro, J. Phys. Conf. Ser. {\bf 143}, 012004 (2009).

\bibitem{heim15}
B. Heim, T. F. R{\o}nnow, S. V. Isakov, and M. Troyer, Science {\bf 348}, 215 (2015).

\bibitem{zanca16}
T. Zanca and G. E. Santoro, Phys. Rev. B {\bf 93}, 224431 (2016).

\bibitem{knysh16}
S. Knysh, Nat. Commun. {\bf 7}, 12370 (2016).

\bibitem{lucas14}
A. Lucas, Front. Phys. {\bf 2}, 5 (2014).

\bibitem{amin18}
M. H. Amin, E. Andriyash, J. Rolfe, B. Kulchytskyy, and R. Melko, Phys. Rev. X {\bf 8}, 021050 (2018).

\bibitem{benedetti16}
M. Benedetti, J. Realpe-G\'omez, R. Biswas, and A. Perdomo-Ortiz, Phys. Rev. A {\bf 94}, 022308 (2016).

\bibitem{altshuler10}
B. Altshuler, H. Krovi, and J. Roland, PNAS {\bf 107}. (28) 12446-12450 (2010);

\bibitem{ronnow14}
T. F. R{\o}nnow, Z. Wang, J. Job, S. Boixo, S. V. Isakov, D. Wecker, J. M. Martinis, D. A. Lidar, and M. Troyer, Science {\bf 345}, 420 (2014).

\bibitem{katzgraber14}
H. G. Katzgraber, F. Hamze, and R. S. Andrist, Phys. Rev. X {\bf 4}, 021008 (2014).

\bibitem{amin15}
M. H. Amin, Phys. Rev. A {\bf 92}, 052323 (2015).

\bibitem{katzgraber15}
H. G. Katzgraber, F. Hamze, Z. Zhu, A. J. Ochoa, and H. Munoz-Bauza, Phys. Rev. X {\bf 5}, 031026 (2015).

\bibitem{mayor15}
V. Martin-Mayor and I. Hen, Sci. Rep. {\bf 5}, 15324 (2015).

\bibitem{venturelli15}
D. Venturelli, S. Mandrà, S. Knysh, B. O’Gorman, R. Biswas, and V. Smelyanskiy, Phys. Rev. X {\bf 5}, 031040 (2015).

\bibitem{marshall16}
J. Marshall, V. Martin-Mayor, and I. Hen, Phys. Rev. A {\bf 94}, 012320 (2016).

\bibitem{marshall17}
J. Marshall, E. G. Rieffel, and I. Hen, Phys. Rev. Applied {\bf 8}, 064025 (2017).

\bibitem{mishra18}
A. Mishra, T. Albash, and D. A. Lidar, Nat. Commun. {\bf 9}, 2917 (2018).

\bibitem{denchev16}
V. S. Denchev, S. Boixo, S. V. Isakov, N. Ding, R. Babbush, V. Smelyanskiy, J. Martinis, and H. Neven, Phys. Rev. X {\bf 6}, 031015 (2016).

\bibitem{chancellor16}
N. Chancellor, G. Aeppli, P. A. Warburton, arXiv:1605.07549.

\bibitem{gardas18}
B. Gardas, J. Dziarmaga, W. H. Zurek, and M. Zwolak, Sci. Rep. {\bf 8}, 4539 (2018).

\bibitem{chengwei15}
C.-W. Liu, A. Polkovnikov, and A. W. Sandvik, Phys. Rev. Lett. {\bf 114}, 147203 (2015).

\bibitem{dwavemanual}
D-Wave User Manual 09-1109A-Q (D-Wave Systems Inc., Burnaby, 2019)

\bibitem{sm}
See Supplementary Material for further discussion of the embedding, KZM scaling in the ordered phase,
additional DWQ data and analysis, TFIM with local noise source, coherence time of a single model spin, and
noise generation in the model.

\bibitem{polkovnikov05}
A. Polkovnikov, Phys. Rev. B {\bf 72}, 161201(R) (2005).

\bibitem{zurek05}
W. H. Zurek, U. Dorner, and P. Zoller, Phys. Rev. Lett. {\bf 95}, 105701 (2005).

\bibitem{dziarmaga10}
J. Dziarmaga, Adv. Phys. {\bf 59}, 1063 (2010).

\bibitem{polkovnikov11}
A. Polkovnikov, K. Sengupta, A. Silva, and M. Vengalattore, Rev. Mod. Phys. {\bf 83}, 863 (2011).

\bibitem{degrandi10}
C. De Grandi, V. Gritsev, and A. Polkovnikov, Phys. Rev. B {\bf 81}, 012303 (2010).

\bibitem{degrandi11}
C. De Grandi, A. Polkovnikov, and A. W. Sandvik, Phys. Rev. B {\bf 84}, 224303 (2011).

\bibitem{degrandi13}
C. D. Grandi, A. Polkovnikov, and A. W. Sandvik, J. Phys.: Condens. Matter {\bf 25}, 404216 (2013).

\bibitem{chengwei13}
C.-W. Liu, A. Polkovnikov, and A. W. Sandvik, Phys. Rev. B {\bf 87}, 174302 (2013).

\bibitem{xu17}
N. Xu, K.-H. Wu, S. J. Rubin, Y.-J. Kao, and A. W. Sandvik, Phys. Rev. E {\bf 96}, 052102 (2017).

\bibitem{xu18}
N. Xu, C. Castelnovo, R. G. Melko, C. Chamon, and A. W. Sandvik, Phys. Rev. B {\bf 97}, 024432 (2018).

\bibitem{kolodrubetz12}
M. Kolodrubetz, B. K. Clark, and D. A. Huse, Phys. Rev. Lett. {\bf 109}, 015701 (2012).

\bibitem{Chandran13}
A. Chandran, F. J. Burnell, V. Khemani, and S. L. Sondhi, J. Phys.: Condens. Matter {\bf 25} 404214 (2013).

\bibitem{Maraga16}
A. Maraga, P. Smacchia, and A. Silva, Phys. Rev. B {\bf 94}, 245122 (2016).

\bibitem{dutta16}
A. Dutta, A. Rahmani, and A. del Campo, Phys. Rev. Lett. {\bf 117}, 080402 (2016).

\bibitem{chenu17}
A. Chenu, M. Beau, J. Cao, and A. del Campo, Phys. Rev. Lett. {\bf 118}, 140403 (2017).

\bibitem{patane08}
D. Patan\`e, A. Silva, L. Amico, R. Fazio, and G. E. Santoro, Phys. Rev. Lett. {\bf 101}, 175701 (2008).

\bibitem{yan18}
Z. Yan, L. Pollet, J. Lou, X. Wang, Y. Chen, and Z. Cai, Phys. Rev. B {\bf 97}, 035148 (2018).

\bibitem{hoyos07}
J. A. Hoyos, C. Kotabage, and T. Vojta, Phys. Rev. Lett. {\bf 99}, 230601 (2007).

\bibitem{hoyos08}
J. A. Hoyos and T. Vojta, Phys. Rev. Lett. {\bf 100}, 240601 (2008).

\bibitem{nishimura16}
K. Nishimura, H. Nishimori, A. J. Ochoa, and H. G. Katzgraber, Phys. Rev. E {\bf 94}, 032105 (2016).

\bibitem{kechedzhi16}
K. Kechedzhi and V. N. Smelyanskiy, Phys. Rev. X {\bf 6}, 021028 (2016).

\bibitem{arceci17}
L. Arceci, S. Barbarino, R. Fazio, and G. E. Santoro, Phys. Rev. B {\bf 96}, 054301 (2017).

\bibitem{keck17}
M. Keck, S. Montangero, G. E. Santoro, R. Fazio, and D. Rossini, New J. Phys. {\bf 19}, 113029 (2017).

\bibitem{patane09}
D. Patan\'e, L. Amico, A. Silva, R. Fazio, and G. E. Santoro, Phys. Rev. B {\bf 80}, 024302 (2009).

\bibitem{nalbach15}
P. Nalbach, S. Vishveshwara, and A. A. Clerk, Phys. Rev. B {\bf 92}, 014306 (2015).

\bibitem{arceci18}
L. Arceci, S. Barbarino, D. Rossini, and G. E. Santoro Phys. Rev. B {\bf 98}, 064307 (2018).

\bibitem{smelyanskiy17}
V. N. Smelyanskiy, D. Venturelli, A. Perdomo-Ortiz, S. Knysh, and M. I. Dykman, Phys. Rev. Lett. {\bf 118}, 066802 (2017).

\bibitem{sun10}
K.-W. Sun, C. Wang, and Q.-H. Chen, Europhys. Lett. {\bf 92}, 24002 (2010).

\end{thebibliography}
\end{document}